\documentclass[a4paper,11pt]{article}

  \usepackage{
    a4wide,
    amsmath,
    amsfonts,
    ascmac,
    bm,
    braket,
    cite,
    comment,
    enumerate,
    here,
    }
  \usepackage[italicdiff]{physics}
  \usepackage{color,graphicx}
  \usepackage[labelsep=quad]{caption}
  \usepackage[labelsep=quad]{subcaption}
  \usepackage{authblk}
  \usepackage{tikz}

\usepackage[normalem]{ulem}

  \usepackage[pdftex]{hyperref}

\newcommand{\figref}[1]{Fig.~\ref{#1}}

\newcommand{\e}[1]{\mathrm{e}^{#1}}

\newcommand{\cst}{{\mathrm{const.}}}

\newcommand{\veps}{\varepsilon}

\definecolor{DarkMagenta}{rgb}{0.54,0,0.54}

\definecolor{DarkBlue}{rgb}{0,0,0.7} 

\definecolor{DarkRed}{rgb}{0.65,0,0}

\begin{document}

\title{
Thermal equilibrium states and instability of self-gravitating particles in an asymptotically AdS spacetime 
}

\author{Hiroki Asami\footnote{Email:asami.hiroki@a.mbox.nagoya-u.ac.jp}}
\author{Chul-Moon Yoo\footnote{Email:yoo@gravity.phys.nagoya-u.ac.jp}}
\affil{Division of Particle and Astrophysical Science, Graduate School of Science, Nagoya University, Nagoya 464-8602, Japan}
\setcounter{Maxaffil}{0}
\renewcommand\Affilfont{\itshape\small}
\date{}

\maketitle

  \begin{abstract}
    We investigate the existence and the stability of spherically symmetric thermal equilibrium states 
    of the self-gravitating many-particle system that satisfies the Einstein-Vlasov equations with a negative cosmological constant.
    While a thermal equilibrium state of the self-gravitating particle system cannot have a finite mass without an artificial wall 
    in the asymptotically flat case, in the asymptotically AdS case, 
    the total mass can be finite due to the AdS potential barrier without any artificial wall. 
    In this case,the AdS radius characterizes the typical size of the system.
    The two independent parameters parametrize the equilibrium states.
    Taking the total rest mass as the unit and fixing the AdS radius, we obtain the 
    one-parameter family of equilibria which describes a curve in the parameter space spanned by the gravothermal energy and the temperature. 
    Then we investigate the instability of the system based on the turning point method for each value of the AdS radius. 
    We find that the curve typically has a double spiral structure as in the asymptotically flat case with an artificial wall. 
    The equilibrium solutions exist only for the finite parameter region of the gravothermal energy between the onsets of the spirals.
  \end{abstract}



  \section{Introduction}
    Self-gravitating systems have distinct thermodynamical properties and have attracted much attention.
    One of the most interesting examples describing such properties is the gravothermal catastrophe~(instability) associated with the negative specific heat.
    Antonov investigated the self-gravitating many-particle system surrounded by a spherical adiabatic wall of the radius $R$
    by introducing a statistical description~\cite{antonov1962}. 
    Ref.~\cite{antonov1962} showed that the Maxwell-J\"{u}ttner distribution gives the distribution function for an equilibrium state, 
    and the system has no global maximum of the entropy while it may have a local maximum.
    They also showed that the second variation of the entropy cannot be negative
    if the ratio between the density at the center and one at the edge of the system is smaller than 709. 
    Lynden-Bell and Wood applied the Poincar\'{e}'s stability criterion\cite{Poincare1885} to the self-gravitating system\cite{BellWood1968}.
    Poincar\'{e}'s criterion says that a series of equilibria can destabilize only when it passes through a turning point. 
    The number of unstable modes changes through a turning point of the series of equilibria~\cite{katz1978,Katz1979,Sorkin1981,Sorkin1982}.
    By using the turning point method, 
    Lynden-Bell and Wood\cite{BellWood1968} showed that no equilibrium solutions exist 
    if the gravothermal energy $E$ is lower than the value $E_\mathrm{min}=-0.335N^2/R$ with $N$ being the number of particles, 
    and the point with this value corresponds to the critical point found by Antonov\cite{antonov1962}.
    In other words, if the radius $R$ of the artificial wall which confines the system is not sufficiently small, the system is unstable. 
    While the above statement is based on the microcanonical ensemble, i.e., the particle number and the total energy are fixed in the variation,
    the thermodynamical stabilities in other processes were studied in Refs.~\cite{BellWood1968, LecarKatz1981, chavanis2002, chavanis2003}.

    In the case of Newtonian gravity, the system can be thermodynamically stable as long as the radius of the artificial wall is sufficiently small. 
    On the other hand, in general relativity, it is expected that the system becomes unstable against gravitational collapse 
    if the confined region is too small compared with the total energy of the system.
    This expectation was shown to be correct in Refs.~\cite{Roupas2014,Roupas:2014sda,Roupas_2015,Roupas2018,Alberti:2019xaj,Chavanis2020} 
    for massive particle systems (see also~\cite{SWZ1981,Chavanis:2007kn} for radiation).
    They showed that a curve for the series of equilibria has a double spiral structure.
    One of the spirals appears in the low energy region, and the other appears in the high energy region. 
    The former one corresponds to the Newtonian gravothermal catastrophe,
    and the latter one, which causes gravitational collapse, is peculiar to general relativity.
    In this paper, we analyze the relativistic many-particle system satisfying the Einstein-Vlasov equations 
    without introducing any artificial wall but with a negative cosmological constant. 

    Apart from the classical gravothermal catastrophe, the asymptotically AdS gravitational system would be attractive in the context of the AdS/CFT 
    correspondence~\cite{Maldacena:1997re,Gubser_1998,witten1998anti} and the AdS instability ~\cite{Bizo__2011}. 
    Because of the confined structure and the nonlinearity of the field equations, 
    arbitrarily small perturbations may form black holes in an asymptotically AdS spacetime. 
    In Ref.~\cite{Moschidis:2017lcr,Moschidis:2017llu}, they showed that arbitrarily small data that form black holes exist for 
    the spherically symmetric Einstein-massless Vlasov system.
    the final states may depend on the symmetry of the spacetime and initial conditions of perturbations~\cite{Dias:2012tq,Maliborski:2013jca,Buchel:2013uba,Balasubramanian:2014cja,Bizon:2014bya,Balasubramanian:2015uua,Dimitrakopoulos:2015pwa,Green:2015dsa,Craps:2014vaa,Craps:2014jwa,Bizon:2015pfa}, 
    and the final fate has not been well understood yet\cite{Choptuik_2018,Masachs_2019}. 
    The self-gravitating particle system considered in this paper might provide a macroscopic model of excitations in the asymptotically AdS spacetime.
    Therefore the thermodynamical stability of the system might be helpful to guess the final fate of the AdS instability. 

    This paper is organized as follows. 
    In Sec.~\ref{sec:eqs}, we derive basic equations and expressions for physical quantities of an equilibrium state 
    of the self-gravitating many-particle system. 
    The parameter dependence of the spatial profile of the system is shown in Sec.~\ref{sec:paraprodep}. 
    The instabilities of the equilibria are discussed based on the turning point method for a typical case 
    in Sec.~\ref{sec:turn}, and we show the parameter dependence of the configuration of the curve 
    describing the series of equilibria in Sec.~\ref{sec:spiral}. 
    Sec.~\ref{sec:conclusion} is devoted to a summary and conclusion. 

    Throughout this paper, we use the geometrized units in which both 
    the speed of light and Newton's gravitational constant are unity, $c=G=1$.

  \section{Equations and physical quantities of an equilibrium state}
\label{sec:eqs}
    Let us consider the relativistic system of self-gravitating massive particles, each of which has the same rest mass. 
    For notational simplicity, hereafter, we use the units in which the rest mass of a particle is unity. 
    Then the number of particles is equivalent to the total rest mass of the system. 
    We focus on a spherically symmetric spacetime, whose metric is given by
      \begin{align}
        \dd{s^2} = -\e{2\nu(r)}\dd{t^2}+\e{2\mu(r)}\dd{r^2}+r^2\dd{\Omega_2^2},
        \label{eq:metric}
      \end{align}
    where $\dd{\Omega_2^2}=\dd{\theta^2}+\sin^2\theta\dd{\phi^2}$ is a line element on the sphere. 
    Considering the one-particle distribution function $f(x^\mu,p^i)$, we can write the energy-momentum tensor of the self-gravitating system as 
      \begin{equation}
      T_{\mu\nu}=\int dV_p p_\mu p_\nu f(x^\mu,p^i), 
      \end{equation}
    where $p^\mu$ is the particle four-momentum and $dV_p$ is the invariant volume element in the momentum space. 
    The invariant volume element in the momentum space can be written as 
      \begin{equation}
      dV_p=2\sqrt{-g}\ d^4p\ \delta(p^2+1)\ \theta(\veps)=\sqrt{-g}\ d^3p/\veps, 
      \end{equation}
    where $\delta$ and $\theta$ are the delta function and the Heaviside's step function, respectively, and $\veps:=-p_t$. 
    We note that, in the last expression, $\veps$ is a dependent variable of $p^r$, $p^\theta$ and $p^\phi$ through the on-shell condition.

    A thermal equilibrium state is realized by the distribution function maximizing the Boltzmann-Gibbs entropy:
      \begin{equation}
        S:=\int_\Sigma \dd{\Sigma_{\mu}}\ s^{\mu}\qc s^{\mu}:=-\int\dd{V_p}p^{\mu}f(\log f-1),
      \end{equation}
    where $\Sigma$ is a $t=\cst$ hypersurface  and $s^{\mu}$ is the entropy current. 
    We introduce the quasi-local mass $M(r)$ as 
      \begin{align}
        M(r) := \int_0^r\dd{s} 4\pi s^2 \rho(s)
      \end{align}
    with $\rho=-T^t_t$. 
    We also define the number of particles $N$ within the radius $r$ as
      \begin{align}
        N(r) = \int_0^r\dd{s} 4\pi s^2n(s),
      \end{align}
    where $n(r):=\e{\nu+\mu}n^{t}$ is the number density with the current $n^\mu$ given by
      \begin{align}
        n^{\mu} := \int dV_p p^\mu f(x^\mu, p^i).
      \end{align}
    Considering isolated systems, we fix the total mass $M:=\lim_{r\to\infty}M(r)$ and the number of particles $N:=\lim_{r\to\infty}N(r)$.
    Then, introducing the positive numbers $\alpha$ and $\beta$ as the Lagrange multipliers,
    we can obtain the distribution function for the thermal equilibrium state from the condition called the Gibbs relation:
      \begin{equation}
        \var{S}+\alpha\var{N}-\beta\var{M}=0.
        \label{eq:eqcond}
      \end{equation}
    As is shown in Refs.~\cite{Ipser1980,SWZ1981}, this condition leads to the Maxwell-J\"{u}ttner distribution function $f=\exp(\alpha-\beta\veps)$. 

    Since the Maxwell-J\"{u}ttner distribution is isotropic in the momentum space, 
    the energy momentum tensor is written in the isotropic form $T^\mu_\nu=\mathrm{diag}(-\rho,p,p,p)$ where $p:=T_r^r$ is the pressure.
    We derive the expression for each component of the energy-momentum tensor 
    according to the method used in the context of the Einstein-Vlasov system\cite{Rein:1998sj,Andreasson:2011ng,Andreasson:2014ina}.
    Introducing the new integration variables $(J,\psi)$ as
      \begin{align}
        p^\theta=:\frac{\sqrt{J}}{r^2}\cos\psi\qc p^\phi\sin\theta=:\frac{\sqrt{J}}{r^2}\sin\psi,
      \end{align}
    we can rewrite the volume element as $dV_p= \frac{\e{\nu+\mu}}{2r^2\veps}dp^r\wedge dJ\wedge d\psi$.
    The relevant regions of $p^r$ and $J$ are $-\infty< p^r<\infty$ and $0\leq J<\infty$. 
    Considering $\veps$ as an independent variable instead of $p^r$, we obtain the relevant region
    $\veps_0:=\e\nu\leq \varepsilon$ and $0\leq J\leq J_{\rm max}:=r^2(\e{-2\nu}\veps^2-1)$.

    By performing the transformation $(p^r,J)\to(\veps,s)$ with $s:=J/J_\mathrm{max}$, we obtain
      \begin{align}
        dV_p=\frac{\veps_0^{-2}}{\sqrt{1-s}}\pqty{\veps^2-\veps_0^2}^{\frac{1}{2}}d\veps\wedge ds\wedge d\psi,
      \end{align}
    where we multiplied the factor two to take into account the degeneracy associated with the sign of $p^r$.
    The energy density and the pressure for the Maxwell-J\"{u}ttner distribution function are given by the integration of $T_t^t$ and $T_r^r$ over the momentum space as
      \begin{subequations}
        \begin{align}
          \rho(r) &= 4\pi\e{\alpha}\bqty{\frac{K_1(\beta\veps_0)}{\beta\veps_0}+\frac{3K_2(\beta\veps_0)}{(\beta\veps_0)^2}},
          \label{eq:rho}\\
          p(r)    &= \frac{4\pi\e{\alpha} K_2(\beta\veps_0)}{(\beta\veps_0)^2},
          \label{eq:pressure}
        \end{align}
        \label{eq:rho_and_p}
      \end{subequations}
      where $K_1$ and $K_2$ are the modified Bessel functions of the second kind.
      The number density for the Maxwell-J\"{u}ttner distribution is also calculated as 
        \begin{align}
          n(r) = 4\pi \e{\alpha+\mu}\frac{K_2(\beta\veps_0)}{\beta\veps_0}\,.
        \end{align}
      The Einstein's equations with a negative cosmological constant $G_{\mu\nu}+\Lambda g_{\mu\nu}=8\pi T_{\mu\nu}$ reduce to the following equations:
        \begin{subequations}
          \begin{align}
            &\e{-2\mu}
            = 1-\frac{2M(r)}{r}+\frac{r^2}{L^2},\\
            &\nu^\prime
            = \frac{\frac{M(r)}{r^2}+\frac{r}{L^2}+4\pi rp(r)}{1-\frac{2M(r)}{r}+\frac{r^2}{L^2}},
            \label{eq:einstein_equation}
          \end{align}
        \end{subequations}
    with the regularity condition at the center: $\mu(0)=0$,
    where $L:=\sqrt{-3/\Lambda}$ is the AdS radius.

    Let us introduce $\rho_c$, $p_c$ and $\nu_c$ as the values of $\rho$, $p$ and $\nu$ at the origin. 
    We also define the following variables at the origin: $w_c=p_c/\rho_c$, 
    $\beta_c:=\beta \e{\nu_c}$, $n_c=n(0)=\beta_cw_c \rho_c$, and $\ell:=(4\pi\rho_c)^{-1/2}$.
    Introducing $x=r/\ell$, $y:=\nu-\nu_c,\ \tilde{\rho}=\rho/\rho_c$ and $\tilde{p}=p/p_c$, 
    we obtain the mass and the particle number as
      \begin{align}
        M(r) = \ell\int_0^x dz z^2\tilde{\rho}(z)\qc N(r) = \ell\int_0^x dz z^2\tilde{n}(z).
      \end{align}
    Then by defining $\tilde{M}(x)=M(r)/l$ and $\tilde{N}(x)=M(r)/l$, we can rewrite 
    Eq.\eqref{eq:einstein_equation} as follows:
      \begin{align}
        \dv{y}{x}=\frac{\frac{\tilde{M}(x)}{x^2}+\frac{x}{\lambda^2}+ w_cx\tilde{p}(x)}{1-\frac{2\tilde{M}(x)}{x}+\frac{x^2}{\lambda^2}}\,.
        \label{eq:ein_dimless}
      \end{align}
    where $\lambda:=L/\ell$ is the normalized AdS radius. 
    We can get a self-consistent equilibrium solution by solving Eq.\eqref{eq:ein_dimless} 
    together with $d\tilde M/dx=x^2 \tilde \rho$. 
    The boundary conditions are given by $y(0)=y^\prime(0)=\tilde M(0)=0$ from the definitions of $y(x)$ and $\tilde{M}(x)$. 
    Giving the parameter set ($\beta_c$,$\lambda$) and solving Eq.\eqref{eq:ein_dimless} from the center to infinity,
    we obtain a unique self-consistent solution.
    In this sense, the equilibrium solution is described by the two independent parameters $\beta_c$ and $\lambda$, 
    therefore it forms a two parameter family of solutions.

    Normalizing the physical quantities with respect to the particle number $N\pqty{r}$, we obtain 
        \begin{subequations}
          \begin{align}
            \tilde{E}(x) := \frac{\tilde{M}(x)-\tilde{N}(x)}{\tilde{N}(x)} &= \frac{\int_0^x \dd{z} z^2\tilde{\rho}(z)}{\beta_cw_c\int_0^x \dd{z} z^2\tilde{n}(z)}-1,\\
            \tilde{L}(x) := \frac{\lambda}{\tilde{N}(x)}   &= \frac{\lambda}{\beta_cw_c\int_0^x \dd{z} z^2\tilde{n}(z)},\\
            \tilde{r}(x) := \frac{x}{\tilde{N}(x)}   &= \frac{x}{\beta_cw_c\int_0^x \dd{z} z^2\tilde{n}(z)}\,,
          \end{align}
        \end{subequations}
    where $\tilde{n}:=n/n_c$ is the normalized particle density.
    We note that we can regard $\tilde E$ and $\tilde L$ at infinity as the independent parameters to specify an equilibrium solution 
    instead of $\beta_c$ and $\lambda$ although there may be a degeneracy in the space of $(\tilde E(\infty),\tilde L(\infty))$. 

  \section{Results}
    \subsection{Parameter dependence of the profile}
      \label{sec:paraprodep}
\begin{figure}[H]
          \begin{subfigure}[]{0.5\linewidth}
            \begin{center}
              \begin{tikzpicture}[inner sep=0pt]
                \node[anchor=south west] (image) at (0,0)
                  {\includegraphics[width=\hsize]{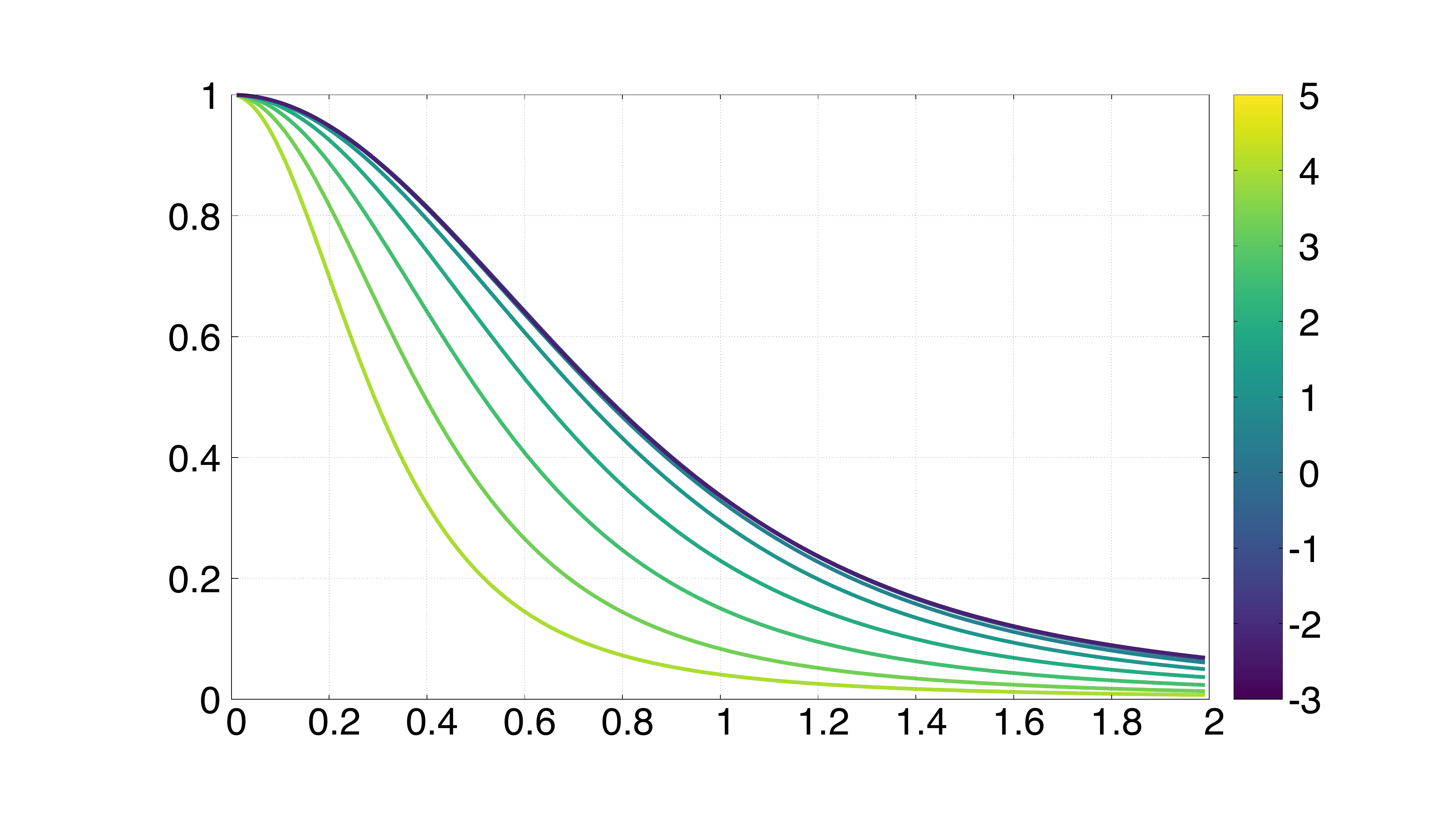}};
                \node at (4.,0) {
                  \scriptsize{$x=r/\ell$}};
                \node at (0.5,2.4) {
                  \scriptsize{\rotatebox{90}{$\tilde{\rho}=\rho/\rho_c$}}};
                \node at (7.4,2.4) {
                  \scriptsize{\rotatebox{90}{$\log\beta_c$}}};
              \end{tikzpicture}
              \subcaption{$\beta_c$ dependence for $\lambda\to\infty$.}
            \end{center}
            \label{fig:DensityProfileflat}
          \end{subfigure}
          \begin{subfigure}[]{0.5\linewidth}
            \centering
            \begin{center}
              \begin{tikzpicture}[inner sep=0pt]
                \node[anchor=south west] (image) at (0,0)
                  {\includegraphics[width=\hsize]{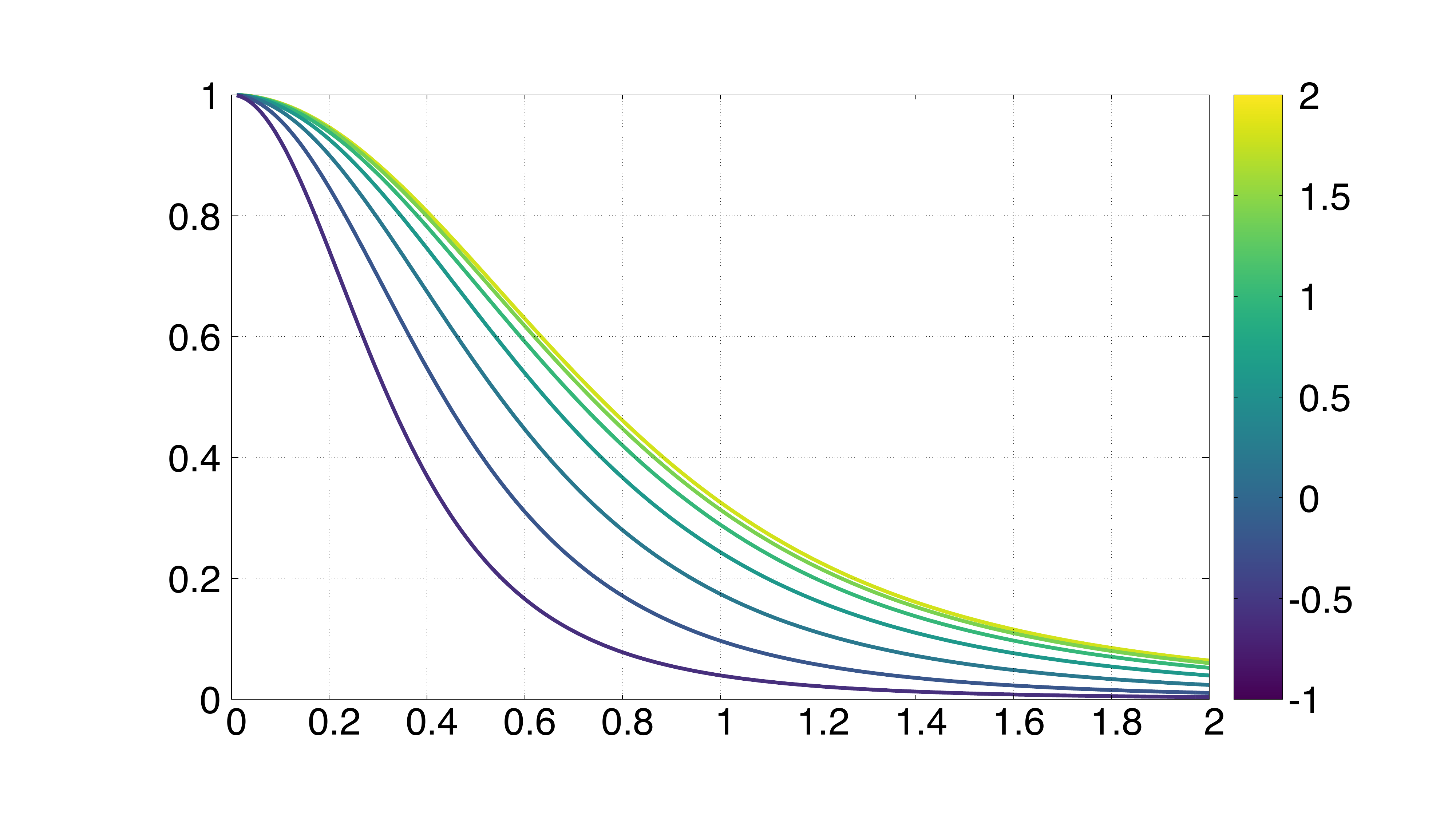}};
                \node at (4.,0) {
                  \scriptsize{$x=r/\ell$}};
                \node at (0.5,2.4) {
                  \scriptsize{\rotatebox{90}{$\tilde{\rho}=\rho/\rho_c$}}};
                \node at (7.4,2.4) {
                  \scriptsize{\rotatebox{90}{$\log\lambda$}}};
              \end{tikzpicture}
                \subcaption{$\lambda$ dependence for $\beta_c=1$.}
            \end{center}
              \label{fig:DensityProfileAdS}
          \end{subfigure}
          \caption{
            The energy density profile as a function of $x$.
            }
          \label{fig:DensityProfile}
        \end{figure}
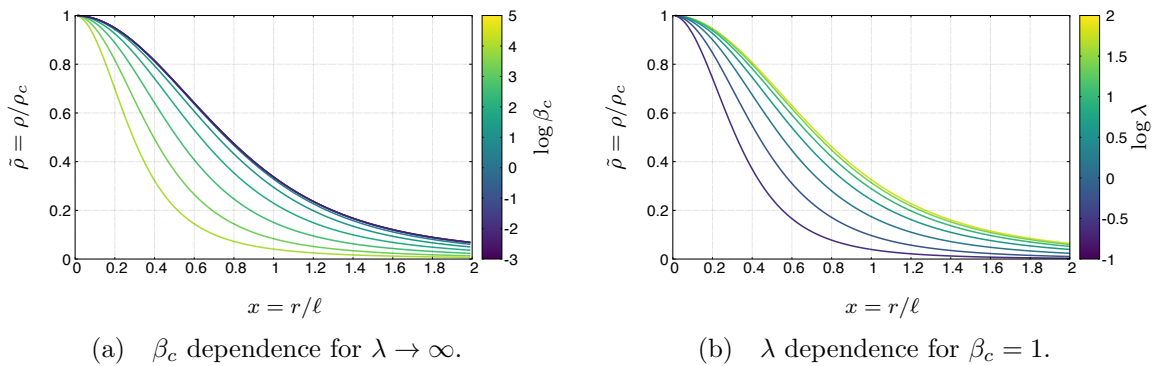

\begin{figure}
          \begin{subfigure}[]{0.5\linewidth}
            \begin{center}
              \begin{tikzpicture}[inner sep=0pt]
                \node[anchor=south west] (image) at (0,0)
                {\includegraphics[width=\hsize]{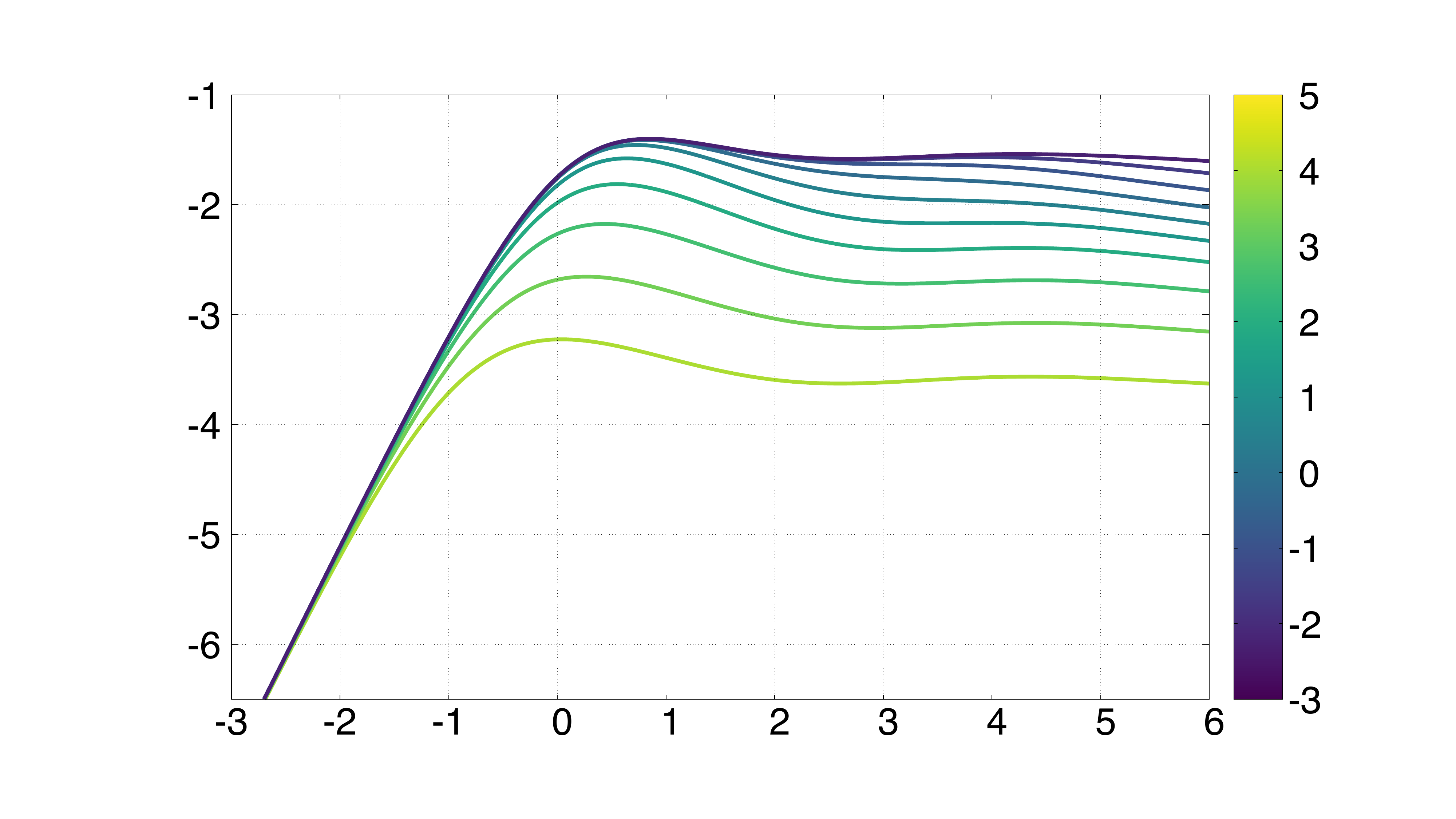}};
                \node at (4.,0) {
                  \scriptsize{$\log x=\log\frac{r}{\ell}$}};
                \node at (0.5,2.4) {
                  \scriptsize{\rotatebox{90}{$\log\frac{M(r)}{r}$}}};
                \node at (7.4,2.4) {
                  \scriptsize{\rotatebox{90}{$\log\beta_c$}}};
              \end{tikzpicture}
            \end{center}
              \subcaption{$\beta_c$ dependence for $\lambda\to\infty$.}
              \label{fig:MassOverRadiusflat}
          \end{subfigure}
          \begin{subfigure}[]{0.5\linewidth}
            \begin{center}
              \begin{tikzpicture}[inner sep=0pt]
                \node[anchor=south west] (image) at (0,0)
                {\includegraphics[width=\hsize]{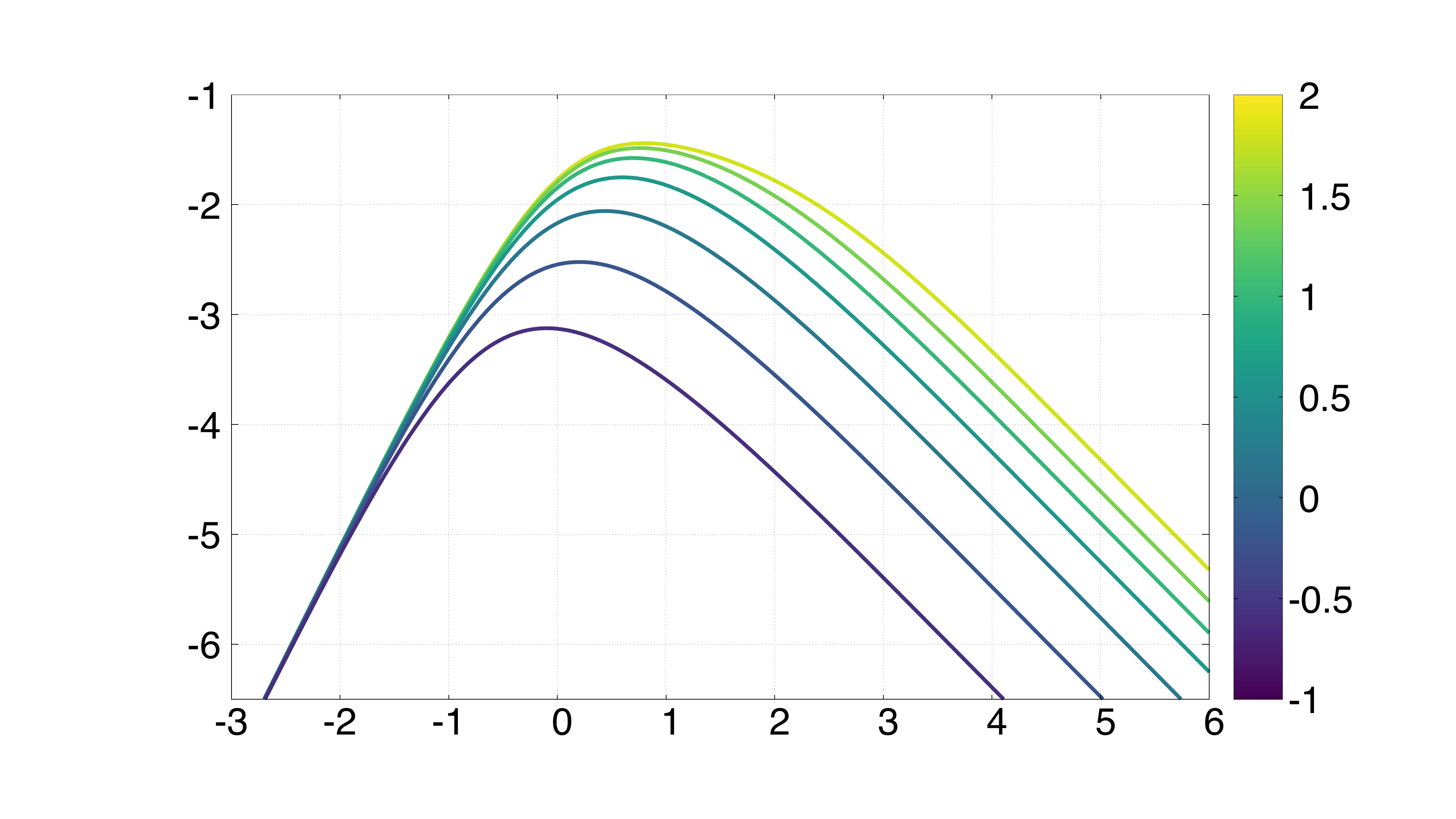}};
                \node at (4.,0) {
                  \scriptsize{$\log x=\log\frac{r}{\ell}$}};
                \node at (0.5,2.4) {
                  \scriptsize{\rotatebox{90}{$\log\frac{M(r)}{r}$}}};
                \node at (7.4,2.4) {
                  \scriptsize{\rotatebox{90}{$\log\lambda$}}}; 
              \end{tikzpicture}
            \end{center}
              \subcaption{$\lambda$ dependence for $\beta_c=1$.}
              \label{fig:MassOverRadiusAdS}
          \end{subfigure}
          \caption{
            The value of $M\pqty{r}/r$ as a function of $x$ in log scale.
            \figref{fig:MassOverRadiusflat} implies that the system limits to the case of the Newtonian gravity 
            in the low temperature limit $\beta_c\to\infty$. 
            \figref{fig:MassOverRadiusAdS} implies that an effective wall exists around $x=\lambda$.}
          \label{fig:MassOverRadius}
        \end{figure}
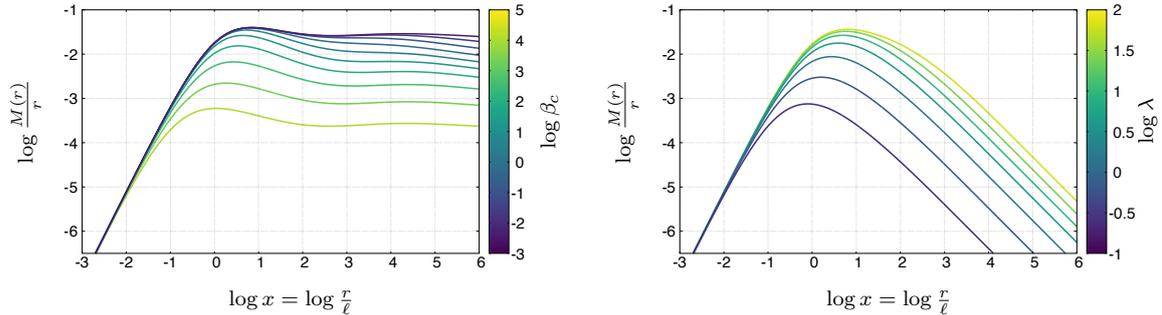

      \figref{fig:DensityProfile} and \figref{fig:MassOverRadius}
      show the density profile and $M(r)/r$ as functions of $r$ for each parameter set $(\beta_c,\lambda)$, respectively.
        \vspace{0pt}%
      As we can see in \figref{fig:DensityProfile} the system becomes more compact as $\lambda$ decreases. 
      We also find $\tilde{\rho}(x)\ll1$ for $x\gg\lambda$ in \figref{fig:DensityProfile},
      and $M\pqty{r}/r\propto 1/x$ for $x\gg \lambda$ for a finite value of $\lambda$ while $M\pqty{r}/r$ asymptotes to some finite value for the case $\lambda=\infty$,
      namely $\Lambda=0$. Therefore the AdS radius $\lambda$ works like the radius of the wall which confines the particles,
      and the total mass $M$ and the particle number $N$ diverge for $\Lambda=0$ without an artificial wall.
      The Newtonian limit can be realized in the low temperature limit, namely $\beta_c\rightarrow \infty$,
      and we find the maximum value of $M\pqty{r}/r$ decreases with increasing $\beta_c$ as is expected in the Newtonian limit.
      Note that the solutions satisfy the Buchdahl inequality $2M(r)/r<8/9$, i.e. $\log{M/r}<log(4/9)\sim-0.81093$.
      This result is consistent with the results in Ref~.\cite{Burikham:2015nma}.

  \subsection{Cold and Hot Turning points}
    \label{sec:turn}
    Before considering our system, let us briefly review the turning point method and the gravothermal catastrophe.
    The self-gravitating system inside the adiabatic wall of the radius $R$ has the instability called gravothermal catastrophe, 
    which can be classified into two types\cite{PADMANABHAN1990285}.
    One is the ``strong instability'' indicating that no equilibrium solutions exist, 
    and the other is the ``weak instability'' indicating 
    that the entropy of the equilibrium is not a local maximum, but just an extremum.
    In the asymptotically flat Newtonian case, the series draws a one-dimensional curve parametrized by 
    $\sigma:=\rho_N(0)/\rho_N(R)$, i.e. $(\bar{E},\bar{\beta})=(\bar{E}(\sigma),\bar{\beta}(\sigma))$, 
    where the variables $\bar E$ and $\bar \beta$ are conjugate with each other in the sense that the variation of the entropy $\bar S$ is given by 
      \begin{equation}
        \delta \bar S=\bar \beta \delta \bar E.  
        \label{eq:varS}
      \end{equation}
    The stability of the system changes at the so-called turning points satisfying $\dv*{\bar{E}}{\sigma}=0$~\cite{Poincare1885}. 
    The turning point corresponds to the point at which one of the eigenvalues of the second variation of the entropy vanishes. 
    Therefore, the weak instability sets in at the turning point.
    Futhermore, the strong instability also sets in at the turning point because the gravothermal energy is bounded by the value at the point.
    It should be noted that, when we apply the turning point method, the thermal equilibrium condition, 
    which we impose on the system, would be essential\cite{katz1978,Katz1979}. 
    Actually, in the Einstein-Vlasov system without thermal equilibrium condition, 
    there is an example indicating the violation of the turning point principle\cite{Gunther:2020mvb}.
    
    In the relativistic system~\cite{Alberti:2019xaj,Chavanis2020}, as is explicitly shown in Sec.~\ref{sec:eqs} for the asymptotically AdS case, 
    the series of equilibria draws a two-dimensional surface.
    In order to apply the turning point method \cite{Poincare1885} to the system, 
    let us consider the one-parameter family of equilibria for each value of $\tilde L$, 
    namely the isolated system with a fixed cosmological constant. 
    Since we take the number of particles as a unit and $\var{N}=0$ in Eq.~\eqref{eq:eqcond},
    the variation of the entropy is given by Eq.~\eqref{eq:varS} with 
    $(\bar{E},\bar{\beta})=(LE/N^2,N\beta/L)\eval_{x\to\infty}$ and $\bar{S}=S/N$.

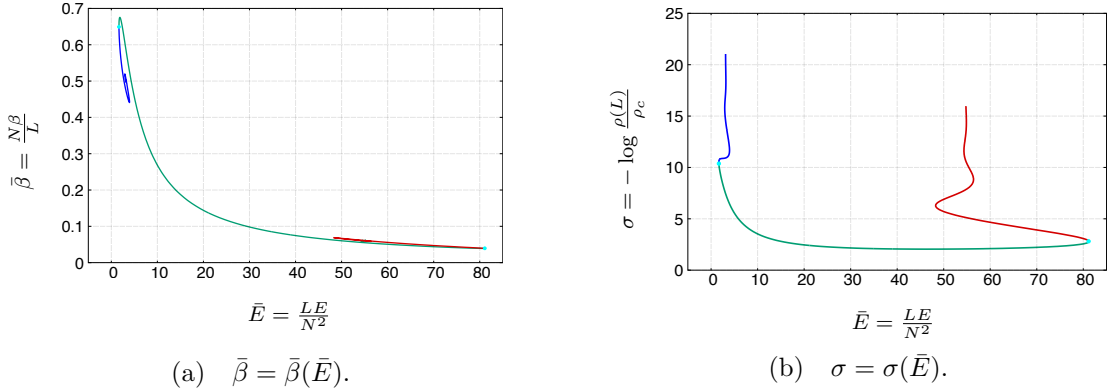
\begin{figure}[htbp]
        \begin{subfigure}[]{0.5\linewidth}
          \centering
            \begin{tikzpicture}[inner sep=0pt]
              \node[anchor=south west] (image) at (0,0)
                {\includegraphics[width=5.8cm]{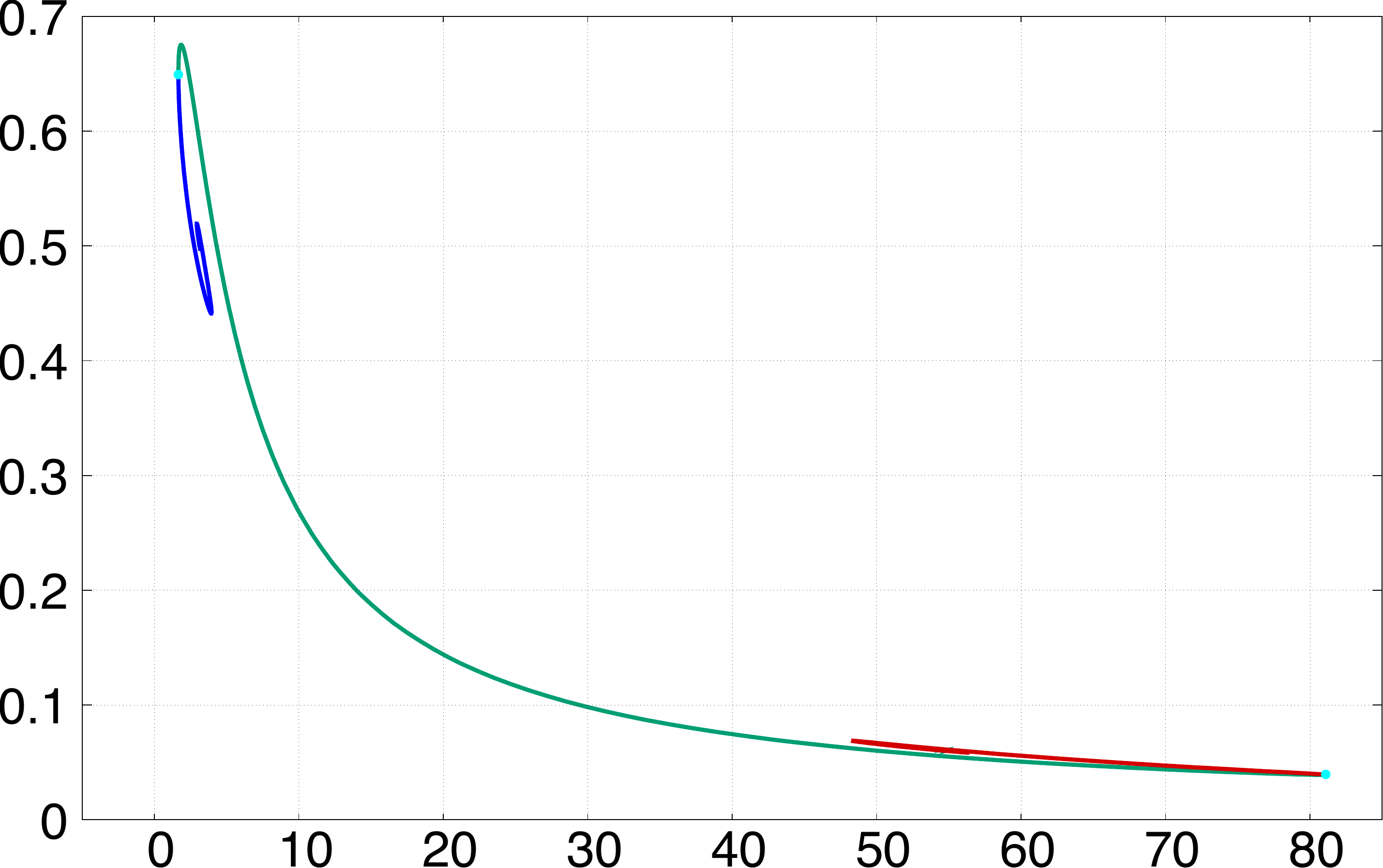}};
              \node at (3,-0.5) {
                \scriptsize{$\bar{E}=\frac{LE}{N^2}$}};
              \node at (-0.5,1.7) {
                \scriptsize{\rotatebox{90}{$\bar{\beta}=\frac{N\beta}{L}$}}};
            \end{tikzpicture}
            \vspace{1ex}
            \subcaption{$\bar{\beta}=\bar{\beta}(\bar{E})$.}
            \label{fig:TypicalCurve}
          \vspace{4ex}
        \end{subfigure}
        \begin{subfigure}[]{0.5\linewidth}
          \centering
            \begin{tikzpicture}[inner sep=0pt]
              \node[anchor=south west] (image) at (0,0)
                {\includegraphics[width=5.8cm]{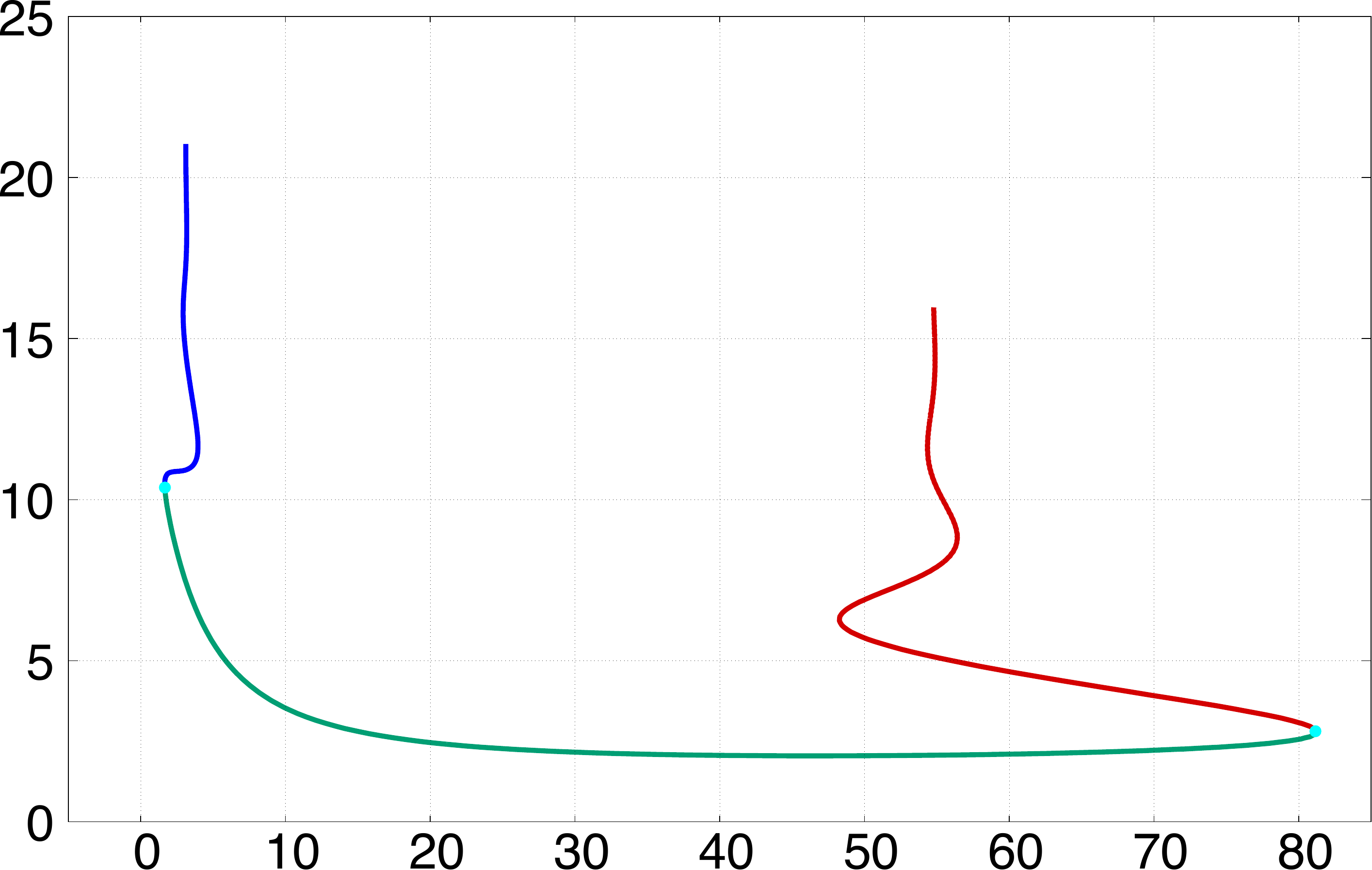}};
              \node at (3,-0.5) {
                \scriptsize{$\bar{E}=\frac{LE}{N^2}$}};
              \node at (-0.5,1.7) {
                \scriptsize{\rotatebox{90}{$\sigma=-\log\frac{\rho(L)}{\rho_c}$}}};
            \end{tikzpicture}
            \subcaption{$\sigma=\sigma(\bar{E})$.}
            \label{fig:DensityCurve}
          \vspace{4ex}
        \end{subfigure}
        \caption{
          \figref{fig:TypicalCurve} shows a typical curve for a series of equilibria with compactness parameter $\gamma = 2.8$.
          \figref{fig:DensityCurve} shows the sharpness parameter $\sigma$ as a function of $\bar E$. 
          The curve has a double spiral structure and we can see that $\sigma$ 
            gets larger from the center to both sides and the curve forms spiral structure after $\sigma$ exceeds a certain value.
          }
        \label{fig:TypicalCurvewithDensityCurve}
      \end{figure}

    An equilibrium state is uniquely identified by a parameter set $(\beta_c,\lambda)$. 
    Then, for a given value of $\gamma:=\log \tilde L$, 
    the sequence is described by a curve on the plane spanned by $\bar E$ and $\bar \beta$ as is shown in \figref{fig:TypicalCurve}. 
    We also define the sharpness parameter $\sigma := -\log\tilde{\rho}(\lambda)$ in analogy with the Newtonian case.  
    \figref{fig:TypicalCurvewithDensityCurve} shows the series of equilibria for $\gamma = 2.8$, 
    and the sharpness parameter is also plotted as a function of $\bar E$.
    A curve for a series of equilibria typically has a double spiral structure.
    This structure in \figref{fig:TypicalCurve} is similar to a curve 
    in a relativistic particle system confined by an artificial wall\cite{Alberti:2019xaj,Chavanis2020}.
    Having this similarity, following Refs.~\cite{Alberti:2019xaj,Chavanis2020},
    we call the right~(high energy) spiral a ``hot spiral'' and the left~(low energy) one a ``cold spiral''.
    We note that, in \figref{fig:DensityCurve}, a curve upward to the right indicates the negative specific heat. 
    Therefore the specific heat, which is the second variation of the entropy $\bar S$ with respect to the gravothermal energy $\bar E$, 
    vanishes at the turning point.
    It means that, along the line towards a spiral, another unstable mode comes in at each turning point on the spiral.
    The sharpness parameter $\sigma$ takes the minimum value at a certain point between the two spirals, 
    and increases towards the end of the spirals as is shown in \figref{fig:TypicalCurvewithDensityCurve}.
    Therefore, towards both ends of the curve, the density profile gets sharper and the system gets more unstable.
    
    The two spirals are characterized by $M/L$ which describes the compactness of the total mass $M$.
    It is related to the gravothermal energy as $M/L=N/L\pqty{N\bar{E}/L+1}=\e{-\gamma}\pqty{1+\e{-\gamma}\bar{E}}$.
    Since the value $\gamma$ is fixed for each curve, the system is more compact for a larger value of $\bar E$.
    Therefore the cold spiral is located in the region where the system is less compact.
    In this case, particles can take energy away from the central region, 
    and the temperature in the central region gets higher because of the negative specific heat. 
    Then the central region gets more compact, and the instability associated with the fragmentation is induced. 
    Therefore the cold spiral corresponds to the well-known Newtonian gravothermal catastrophe.
    In contrast with the cold spiral, the hot spiral is formed because the system is too compact to be stable. 
    That is, the system collapses if too many particles and too much energy are forced to cram into a small region. 
    Such a behavior can be interpreted by considering the effective potential for a particle motion in a Schwarzschild-AdS spacetime.
    When the AdS radius is sufficiently large, the effective potential has a local minimum and a stable orbit exists.
    On the other hand, when the AdS radius is smaller than a certain value, the potential minimum and the stable orbit disappear,
    and the particles inevitably fall into the center of the system.

  \subsection{Parameter dependence of the spiral structure}
  \label{sec:spiral}
\begin{figure}[H]
        \begin{subfigure}[]{0.5\linewidth}
          \centering
            \begin{tikzpicture}[inner sep=0pt]
              \node[anchor=south west] (image) at (0,0)
                {\includegraphics[width=8cm]{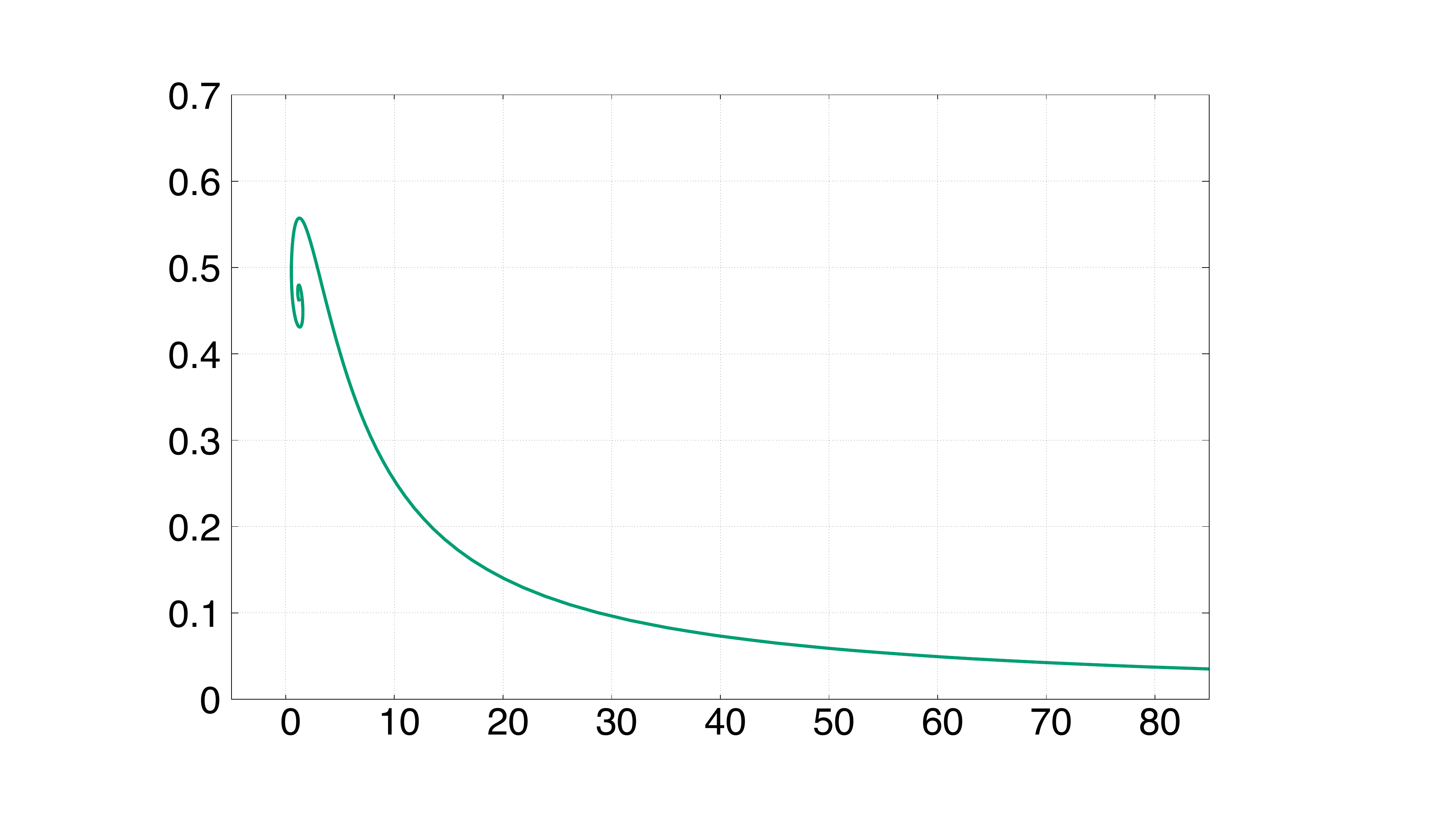}};
              \node at (4.,0) {
                \scriptsize{$\bar{E}=\frac{LE}{N^2}$}};
              \node at (0.5,2.5) {
                \scriptsize{\rotatebox{90}{$\bar{\beta}=\frac{N\beta}{L}$}}};
            \end{tikzpicture}
            \vspace{0.5ex}
            \subcaption{$\gamma=4.00$.}
            \label{fig:LargeGamma}
        \end{subfigure}
        \begin{subfigure}[]{0.5\linewidth}
          \centering
            \begin{tikzpicture}[inner sep=0pt]
              \node[anchor=south west] (image) at (0,0)
                {\includegraphics[width=8cm]{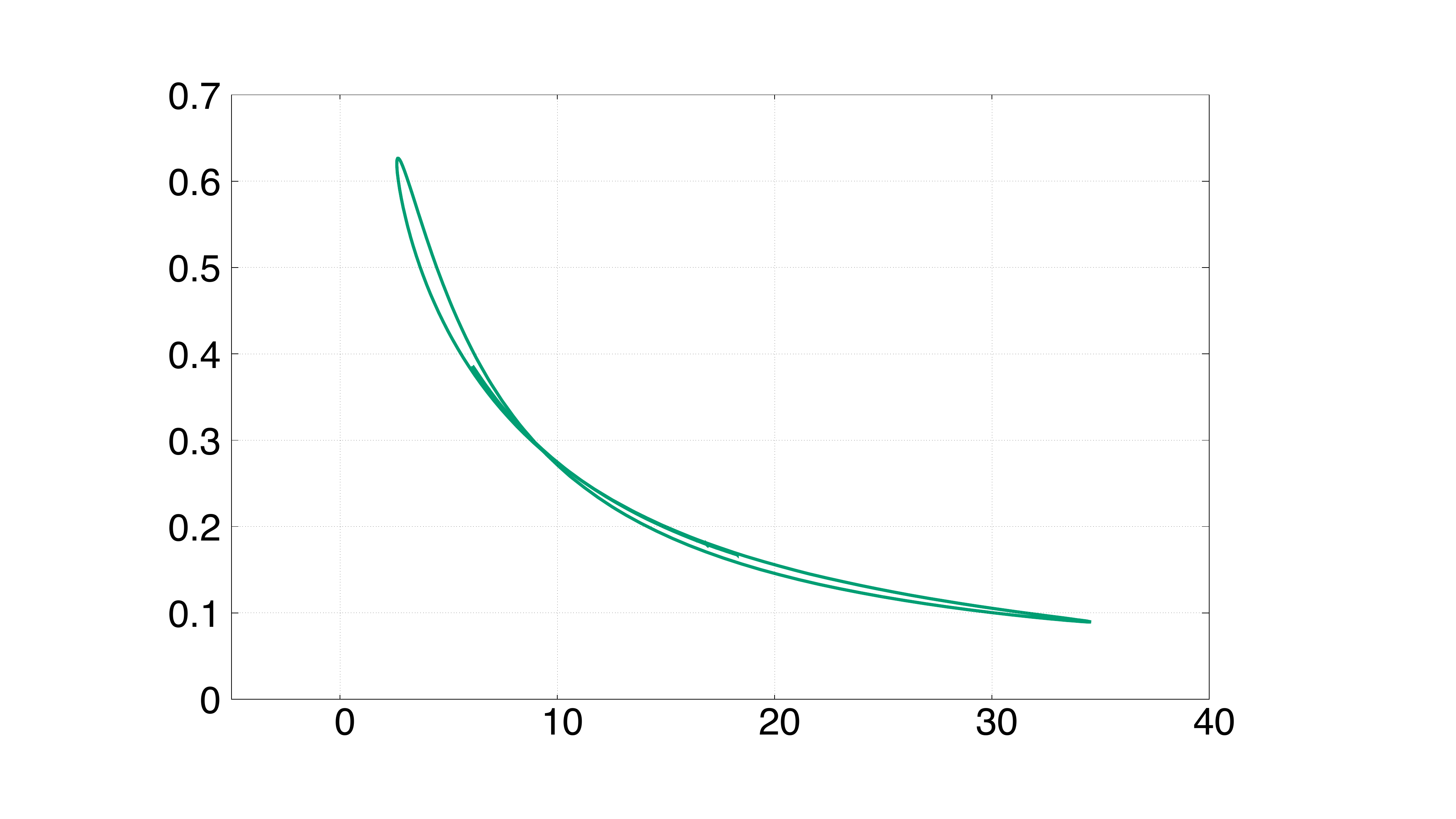}};
              \node at (4.,0) {
                \scriptsize{$\bar{E}=\frac{LE}{N^2}$}};
              \node at (0.5,2.5) {
                \scriptsize{\rotatebox{90}{$\bar{\beta}=\frac{N\beta}{L}$}}};
            \end{tikzpicture}
            \vspace{0.5ex}
            \subcaption{$\gamma=2.50$.}
            \label{fig:CaloricCurveGamma30}
        \end{subfigure}
        \begin{subfigure}[]{0.5\linewidth}
          \centering
            \begin{tikzpicture}[inner sep=0pt]
              \node[anchor=south west] (image) at (0,0)
                {\includegraphics[width=8cm]{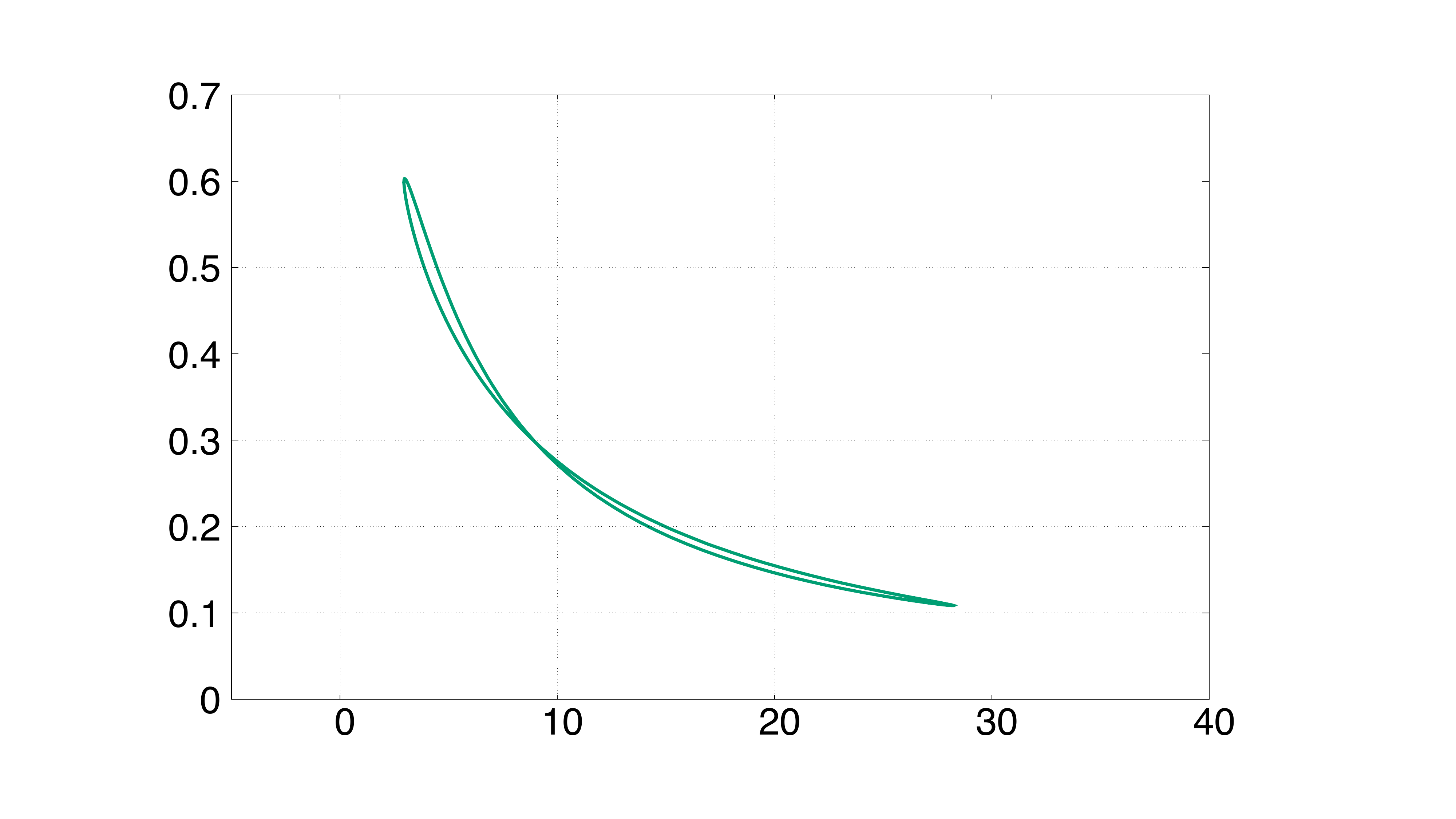}};
              \node at (4.,0) {
                \scriptsize{$\bar{E}=\frac{LE}{N^2}$}};
              \node at (0.5,2.5) {
                \scriptsize{\rotatebox{90}{$\bar{\beta}=\frac{N\beta}{L}$}}};
            \end{tikzpicture}
            \vspace{0.5ex}
            \subcaption{$\gamma=2.44$.}
            \label{fig:CaloricCurveGamma29}
        \end{subfigure}
        \begin{subfigure}[]{0.5\linewidth}
          \centering
            \begin{tikzpicture}[inner sep=0pt]
              \node[anchor=south west] (image) at (0,0)
                {\includegraphics[width=8cm]{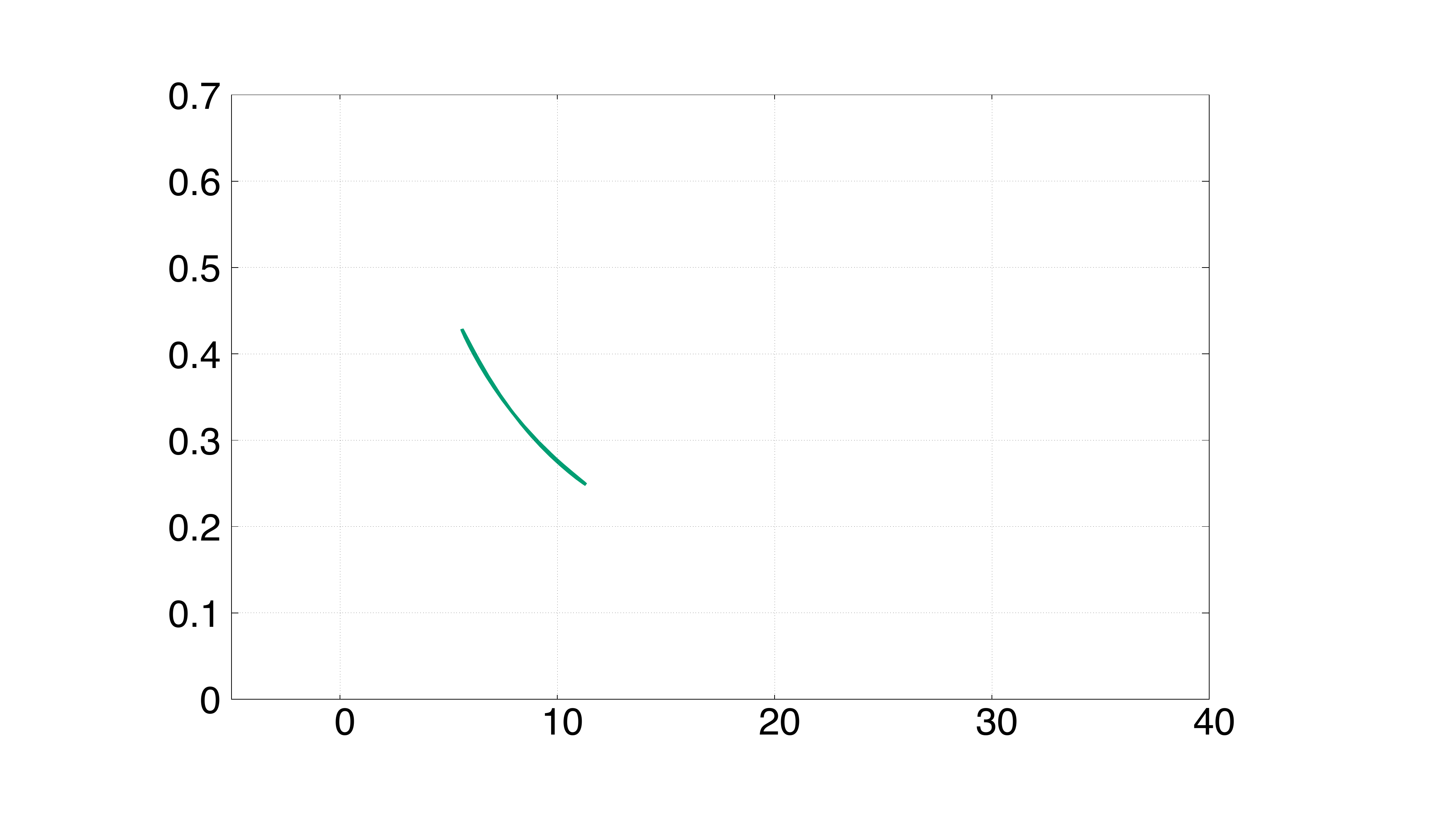}};
              \node at (4,0) {
                \scriptsize{$\bar{E}=\frac{LE}{N^2}$}};
              \node at (0.5,2.5) {
                \scriptsize{\rotatebox{90}{$\bar{\beta}=\frac{N\beta}{L}$}}};
            \end{tikzpicture}
            \vspace{0.5ex}
            \subcaption{$\gamma=2.25$.}
            \label{fig:CaloricCurveGamma32}
        \end{subfigure}
        \vspace{1ex}
        \caption{
          The series of equilibria for various $\gamma=\log\frac{L}{N}$.
          There are three critical points $\gamma_\mathrm{m}, \gamma_\mathrm{\ell}, \gamma_\mathrm{v}$ and
          there are no equilibrium solutions for $\gamma<\gamma_\mathrm{v}$.}
        \label{fig:CaloricCurveVariousGamma}
      \end{figure}
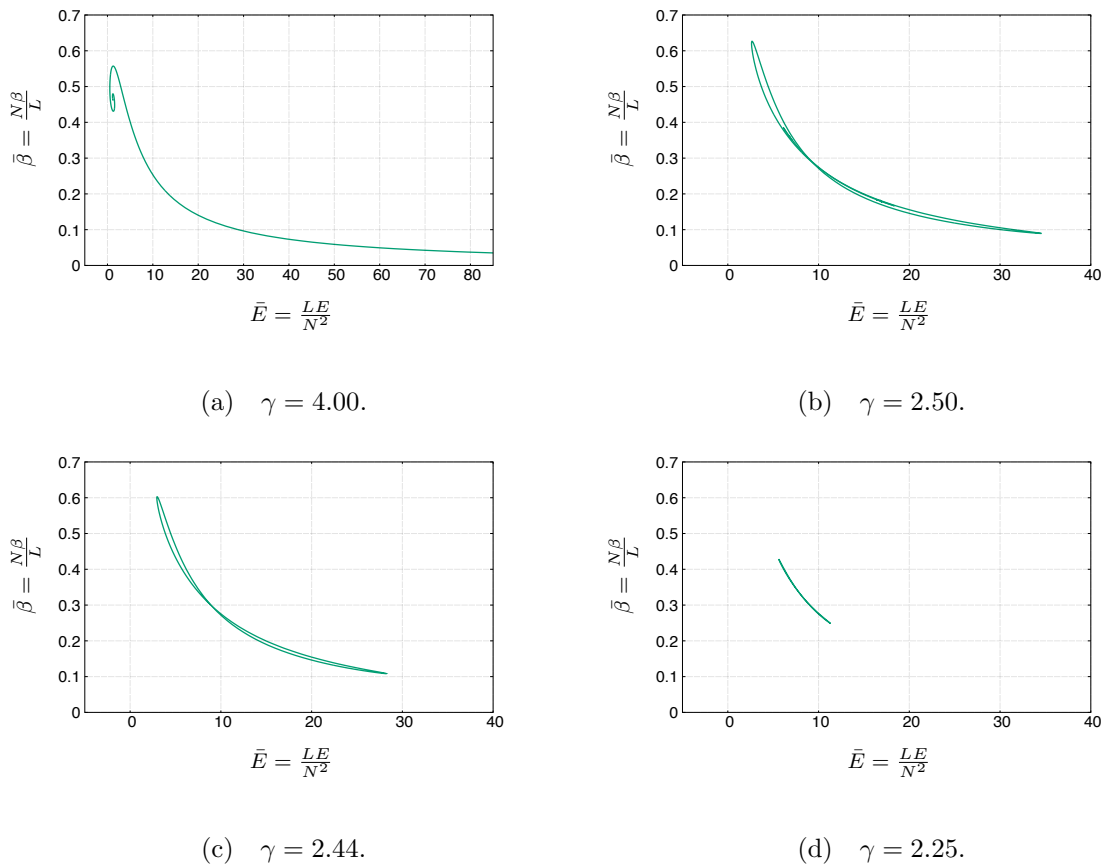

    Let us consider the $\gamma$-dependence of the configuration of the curve.
    If we take $\gamma\to\infty$, namely in the dilute limit, 
    the hot spiral moves to the right infinitely, which corresponds to the Newtonian limit.
    As in the case of the relativistic system with an artificial wall, two spirals approach to each other as $\gamma$ gets smaller,
    and they merge at the ``merge point" $\gamma=\gamma_\mathrm{m}\simeq2.5300$.
    As $\gamma$ gets even smaller, 
    two spirals connect to each other and form a loop at the ``loop point" $\gamma=\gamma_\mathrm{\ell}\simeq2.4468$.
    If we continue to make $\gamma$ smaller, the loop shrinks and vanishes at the ``vanishing point'' $\gamma=\gamma_\mathrm{v}\simeq2.2286$.
    All these behaviors are shown in \figref{fig:CaloricCurveVariousGamma} by plotting curves 
    for some specific values of $\gamma$, and qualitatively similar to the case of the relativistic system 
	with an artificial wall reported in Refs.~\cite{Alberti:2019xaj,Chavanis2020}.

  \section{Conclusion}
\label{sec:conclusion}
    We have studied the existence and the stability of the thermal equilibrium states of 
	a self-gravitating many-particle system confined
    by an AdS potential by applying the turning point method to the series of one-parameter families of the equilibria. 
    We found that all properties, which are explained below, are similar to the case of the relativistic system
    confined with an artificial wall without a cosmological constant but confined with an artificial wall~\cite{Alberti:2019xaj,Chavanis2020}.
    In our analyses, we regard the AdS radius as the size of an effective wall.

    The equilibria can be parametrized by two parameters.
    Fixing the rest mass compactness $\gamma$, 
    the set of equilibria reduces to a one-dimensional curve in 
    the space spanned by the normalized gravothermal energy $\bar E$ and the inverse temperature $\bar\beta$.
    Each curve typically has a double spiral structure indicating two independent instabilities.
    One of them, called a cold spiral, appears at the region of the lower gravothermal energy $\bar E$.
    The other spiral called a hot spiral appears in the region of the higher gravothermal energy.
    Therefore, the equilibrium solutions exist only for the finite parameter region of the gravothermal energy between the onsets of the two spirals, 
    and it indicates the strong instability.
    The turning points are also onsets of weak instabilities.
    The cold spiral corresponds to the well-known Newtonian gravothermal catastrophe while the hot spiral corresponds
    to the instability associated with the strong gravity of the relativistic compact object. 
    Although the origins of the instabilities are different, both of them are associated with the negative specific heat.

    The configuration of the curve in the $(\bar{E},\bar{\beta})$ space depends on the value of $\gamma$.
    In the limit $\gamma\to\infty$, the hot spiral moves away to infinity,
    and there are three critical values at which the topology of the curve changes.
    At the merge point $\gamma_\mathrm{m}\simeq2.5300$, two spirals merge with each other.
    At the loop point $\gamma_\mathrm{\ell}\simeq2.4468$, the curve forms a loop and the two endpoints disappear.
    At the vanishing point $\gamma_\mathrm{v}\simeq2.2286$, the loop vanishes and no equilibrium solutions exist for $\gamma<\gamma_\mathrm{v}$.
    Thus the system cannot be a thermal equilibrium state if too many particles are confined in a small region.
    
    It should be noted that we have not shown that the solutions between two spirals are stable.
    In the Newtonian system with an artificial adiabatic wall,
    the stability of the system can be analytically proven by showing all eigenvalues of the second variation of 
    the entropy are negative if the equilibrium state is not in the cold spiral.
    The same analysis cannot be directly applied to our case, and the clarification of the stability is left as a future work.
    
  \section*{Acknowledgments}
    We would like to thank T. Harada for useful discussions.
    This work was supported by JSPS KAKENHI Grant Numbers JP19H01895~(C.Y.), JP20H05850~(C.Y.), JP20H05853~(C.Y.). 

\appendix
  \def\thesection{Appendix\Alph{section}}
  \renewcommand{\theequation}{A.\arabic{equation}}
  \setcounter{equation}{0}

\bibliographystyle{unsrt}
  \bibliography{citation.bib}

\end{document}